\documentclass{article}
\usepackage{eleco2013,amssymb,amsmath,epsfig}
\usepackage[para,online,flushleft]{threeparttable}
\usepackage{multirow}
\usepackage{dirtytalk}
\usepackage{booktabs}
\usepackage[latin5]{inputenc}
\usepackage{indentfirst}
\usepackage{caption}
\usepackage{mwe} % to get dummy images

  \ninept

\title{\bf An Inception Inspired Deep Network to Analyse Fundus Images}
\name{Fatmatulzehra Uslu$^1$}
\address{$^1$ Bursa Technical University, Electrical-Electronic Engineering Department, Bursa, Turkey\\ {fatmatulzehra.uslu@btu.edu.tr}
 }

\begin{document}

\maketitle
\ninept

\begin{abstract}
 \noindent \bf{A fundus image usually contains the optic disc, pathologies and other structures in addition to vessels to be segmented. This study proposes a deep network for vessel segmentation, whose architecture is inspired by inception modules. The network contains three sub-networks, each with a different filter size, which are connected in the last layer of the proposed network. According to experiments conducted in the DRIVE and IOSTAR, the performance of our network is found to be better than or comparable to that of the previous methods. We also observe that the sub-networks pay attention to different parts of an input image when producing an output map in the last layer of the proposed network; though, training of the proposed network is not constrained for this purpose. }
%This is a great paper and it has a concise abstract. max page number is $8$, excluding references and acknowledgements. It may be extended if it is necessary.
\end{abstract}

%\begin{keywords}
%Inception networks, U-net, retinal vessel segmentation, attention maps  %List of keywords, comma separated.
%\end{keywords}

\section{Introduction}
Deep learning methods have been shown to produce state of the art performance for many image analysis problems \cite{litjens2017survey}. Their performance can be associated with the use of deeper networks with very large number of parameters, residual connections \cite{he2016deep} and regularisation techniques such as dropout \cite{srivastava2014dropout}. However, the performance of medical image analysis problems is limited by amounts of available labelled data. %so effective use of data is a necessity.

The segmentation of retinal vessels in fundus images may be required to extract topology or to measure bifurcation angles or vessel widths \cite{fraz2012blood,uslu2019recursive,uslu2018multi}. Although vessels appear to be darker than surroundings in a fundus image, they may be confused with pathologies or the border of the optic disc or the field of view (FOV). Also, the central light reflex, a bright stripe running along a vessel centreline, may cause segmentation methods to perceive it as background between two thin vessels \cite{fraz2012blood}. 

Many studies using deep networks for vessel segmentation have trained a single network to label the location of vessels in a pixel-wise manner \cite{liskowski2016segmenting,Li2015}. However, when the image background is cluttered or there is large variations on the appearance of vessels, the capacity of a deep network, which is often limited by the amount of labelled data, may need to be used for eliminating uneven illumination, large noise and other structures such as the optic disc and pathologies, in addition to precisely finding the location of vessels.

%Fundus cameras and the scanning laser opthalmoscopes are the most used imaging modalities to capture the back of the eye \cite{yannuzzi2004ophthalmic}
%Deep learning methods have been shown to produce better performance metrics at many medical image analysis problems\cite{litjens2017survey}. However, the application of deep learning methods to medical problems usually requires high understanding about the decision given by the methods. Until recently, deep learning methods were called "black box" because it was difficult to interpret their internal decision mechanism. Recently, some methods have been proposed to make a better understanding on which grounds networks make their decisions. The methods have studied ways to visualise activation maps or feature sets generated at any convolutional layer of a previously trained deep network for specific sets of input images  \cite{yosinski2015understanding,zhou2018interpreting,esteva2017dermatologist}. Also, attention maps have been used to show and explain where the attention of a network is focused for a specific task \cite{wang2017zoom,guan2018diagnose,oktay2018attention}. 

This study presents a deep network for vessel segmentation in fundus images. The architecture of the network is inspired by inception modules \cite{szegedy2015going}. The proposed network consists of three sub-networks, each with a different filter size. The decision given by each sub-network is combined in the last layer of the proposed network to produce a single decision. %The sub-networks share the same loss layer and are simultaneously trained. 
We evaluate the performance of the proposed network on two fundus image datasets generated by different imaging modalities: DRIVE and IOSTAR datasets. Based on our experiments, we observe that each sub-network seems to specialise at a different region of an input image, without giving any supervision for it: pixels inside FOV versus those outside FOV, non-vessel pixels inside FOV and vessel pixels inside FOV.

\section{Related Work}

\subsection{U-net}

U-net is one of the best known architectures in medical image segmentation \cite{ronneberger2015u}. The architecture has two paths: one for encoding the input image and the other one for decoding the corresponding segmentation map. Two paths were connected with skip connections to improve gradient flow through the network. Skip connections also provide access to features produced at early layers in generation of the segmentation maps due to the concatenation of the features at encoding path and those at decoding path. In the architecture, there are nine convolutional layers. Each covolutional layer contains two $3 \times 3$ filters. The filters are followed by pooling along the encoding path to half the grid size. They are preceded by upsampling along the decoding path to double the grid size.

%U-net is a fully convolutional network with encoder and decoder parts, which are connected with skip connections \cite{ronneberger2015u}. The network consists of five blocks of two convolutional layers followed by a pooling layer in the encoder part and the decoder part has the symmetric architecture of the encoder and up-sampling layers replace pooling layers to increase the resolution of feature maps. Typically, the filter numbers are increased by a factor of $2$ at each down-sampling    layer and reduced by the same rate at each up-sampling layer. %Each module in encoder part reduce the resolution of input features by half. In contrast, each module in decoder part increases the resolution of the features by a factor of $2$. Skip connections between encoder and decoder modules facilitate the network to access early layer features while generating image masks, which helps to recover details in images lost due to pooling operations. Figure \ref{fig:Unet} depicts the architecture of the U-net .

\subsection{Inception Modules}
\begin{figure}[!t]
 \centering
  {\includegraphics[scale=0.15]{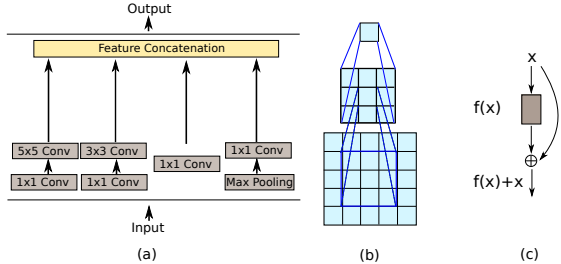}  }
  \caption{(a) The inception network with pooling \cite{szegedy2015going,szegedy2017inception} (b) Realisation of $5 \times 5$ filter with two $3 \times 3$ filters (c) A convolutional layer with a residual connection. (Best viewed in color.) \label{fig:InceptionModels}}
%\vspace{-0.25cm}
\end{figure}

Inception modules were proposed by Szegedy \textit{et al.} \cite{szegedy2015going,szegedy2017inception}, which were used in \say{GoogleNet}. The inception modules were designed to be micro-networks \cite{szegedy2015going}, which can be located at desired depths of a macro-network. The main characteristic of inception modules is that they contain a range of filter types in parallel; a basic version consists of $1 \times 1$, $3 \times 3$, $5 \times 5$ filters. This structure provides rich feature set for the next convolutional layer of a deep network by leading to better performance with a small parameter number, which was proven by the performance of "GoogleNet" \cite{szegedy2015going} in ImageNet Large-Scale Visual Recognition Challenge $2014$ (ILSVRC$14$). In order to reduce the parameter number of the inception modules, one may use $1 \times 1$ filters to decrease the channel number, as illustrated in \textbf{Fig.} \ref{fig:InceptionModels}(a), or replace large filters with small ones; a $5 \times 5$ filter is factorised to two $3 \times 3$ filters in \textbf{Fig.} \ref{fig:InceptionModels}(b).

\subsection{Residual Connections}
He \textit{et al.} integrated residual connections into deep networks and showed an increase in the performance of the networks due to the residual connections facilitating better gradient flow through the layers of the networks \cite{he2016deep}. As shown in \textbf{Fig.} \ref{fig:InceptionModels}(c), the output of a convolutional layer with a residual connection becomes the sum of the output of the convolutional layer ($f(x)$) and its input ($x$).

\section{Method}
Inspired by the architecture of the inception modules, we present a network with three parallel sub-networks, each with a different filter size. The outputs of the sub-networks are combined with a final convolutional layer to allow them both to be jointly trained and to produce a single vessel mask for an input image. We expect the sub-networks to act as three experts giving a joint decision on the generation of vessel masks.

\begin{figure*}[!t]
 \centering
  {\includegraphics[scale=0.45]{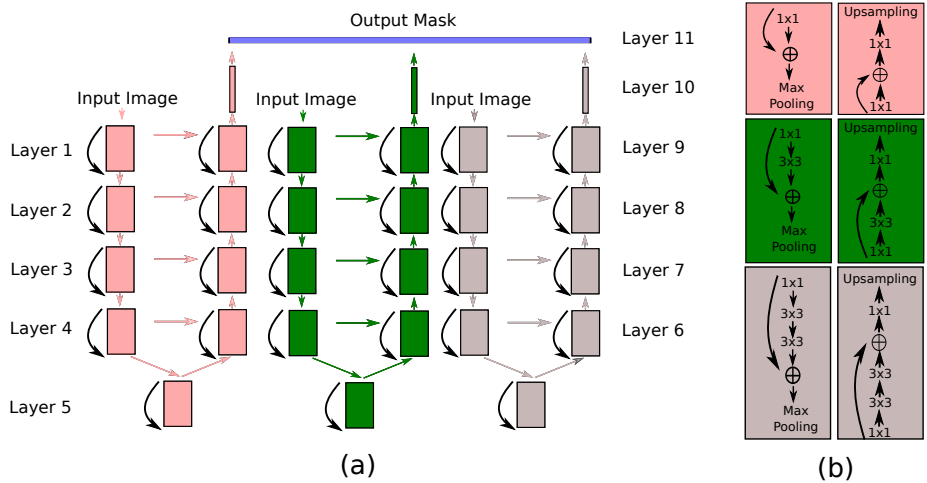}  }
  \caption{(a)An overview of the proposed architecture. Pink, green and brown boxes respectively show modules with $1 \times 1$, $3 \times 3$ and $5 \times 5$ filters. Pink, green and brown arrows respectively demonstrate the connection between filters with the same colour code as the arrows and black arrows show residual connections. Blue box illustrates the final convolutional layer of the proposed network. Narrow boxes in pink, green and brown represent filters with $1$ channel. The numbers of channels in the other filters are not associated with the widths of the boxes. Input images for each sub-network is the same. (b) Filters inside modules with the same colur code. (Best viewed in color.) \label{fig:Method}}
%\vspace{-0.25cm}
\end{figure*}

\textbf{Fig.} \ref{fig:Method} illustrates the general structure of the proposed architecture. We use U-net as a sub-network. Each sub-network contains modules with one of predetermined filter types, which are $1\times 1$, $3 \times 3$ and $5 \times 5$ filters.  %Filter types used in the sub-networks are $1\times 1$, $3 \times 3$ and $5 \times 5$. 
In order to increase gradient flow through the network, we integrate residual connections into the modules \cite{he2016deep}. In contrast to U-net, we use one module, in the place of two convolution layers \cite{ronneberger2015u}, at each grid size and padded features prior to applying filters to preserve their size. In the final layer of each sub-network, we reduce the number of filters to $1$ by using $1 \times 1 $ convolution, which can be viewed as each sub-network producing its own object mask. Eventually, three features generated by the sub-networks are combined with $1 \times 1 $ convolution on the top layer of the proposed network to produce a single object mask. We train the network as a regressor, with Euclidean loss between vessel masks synthesised by the network and ground truth images.

Similar to the U-net \cite{ronneberger2015u}, we double filter sizes along the encoding path of the proposed network and halve them along its decoding path. When downsampling features along the encoding path, we use $2 \times 2$ max pooling with a stride of $2$ for the sub-network with a $1 \times 1$ filter and apply convolution with a stride of $2$ pixels for the other two sub-networks. When upsampling features along the decoding path, we use bilinear interpolation after reducing channel number of incoming features by a factor of $2$ with $1 \times 1$ filters to lighten computation burden, in contrast to U-net \cite{ronneberger2015u} where channel reduction was performed after upsampling.

Each convolutional layer of our network is preceded by batch normalisation and followed by RELU activation function, apart from the one in layer $11$, which is followed by sigmoid function. Apart from the modules with  $1 \times 1$ filters, we use $1 \times 1$ filters in each module regardless of its filter type to reduce the number of incoming filters, prior to applying filters specific to each sub-network. Similar to inception modules \cite{szegedy2015going,szegedy2017inception},  we factorise $5 \times 5$ filters to two $3 \times 3$ filters to reduce the parameter number of the proposed network, which is $250,933$ in total. \textbf{Table} \ref{tab:ChannelNo1x1} shows the number of filters for each layer of sub-networks regardless of filter type. %We used the same number of filters for two $3 \times 3$ filters in the sub-network with $5 \times 5$ filters, which is equal to the number of $3 \times 3$ filters in \textbf{Table} \ref{tab:ChannelNo3x3}. 

  %\textbf{(when filter number was halved, the total number of parameters reduce to $\mbox{250,813}$.) } (\textbf{ that of one with residual links}). 

%\textbf{When I used ELU function instead of RELU and BN, I keep RELU and BN in layer $10$ and changed the rest.}
%We also add a  residual link for each filter type. 

%use ELU and transposed convolutional layer for upsampling. 

\begin{table}[t] 
\centering
\caption{Filter numbers for the layers of the sub-network with $1 \times 1$ filters.}
\label{tab:ChannelNo1x1}
\resizebox{0.3\textwidth}{!}{%
%\begin{threeparttable}
\begin{tabular}{ @{} l c  c  @{}}  \toprule
Layer No & Filter No & Filter No in Upsampling Layer \\  \cmidrule(lr){1-1}  \cmidrule(lr){2-2} \cmidrule(lr){3-3} 
1 & 8 &  \\
  2 & 16 & \\
  3 & 32&  \\
  4 & 64 &  \\
  5& 128& 64 \\
  6& 64 & 32 \\
  7& 32& 16\\
   8& 16& 8 \\
   9& 8 & \\
   10& 1 & \\ \bottomrule								
\end{tabular}
}
\end{table}

\section{Material and Experimental Setup}
\subsection{Material}
The segmentation performance of the proposed network was assessed on a well known fundus image dataset, the DRIVE \footnote{https://www.isi.uu.nl/Research/Databases/DRIVE/} \cite{staal2004ridge} and a recently released fundus image dataset, IOSTAR\footnote{http://www.retinacheck.org/datasets}. The DRIVE was captured by Canon CR$5$ with FOV of $45^{o}$ from generally healthy people. The number of images in the dataset is $40$; of these, $7$ show the signs of diabetic retinopathy. The resolution of images is $768 \times 584$ pixels. In order to allow a fair performance comparison, the dataset is divided into two: one for training and the other one for testing. Each set contains $20$ images. Manually traced vessel maps and FOV masks are also provided with the dataset. 

IOSTAR dataset  \cite{zhang2016robust} contains $30$ Scanning Laser Ophthalmoscopy (SLO) images, which were captured by an EasyScan camera with a FOV of $45^{o}$. The resolution of images is $1024 \times 1024$ pixels. FOV masks and binary vessel masks are included in the dataset. We used the first $20$ images as the training and validation sets and evaluated the performance of the proposed method on the the last $10$ images. 

\subsection{Experimental Setup \label{sec:ExperimentalSetup}}

We initially set learning rate to $0.0008$ then decreased it with an exponential decay rate of $0.94$ during training. %. The learning rate for an epoch ($n=N$) was calculated with  $\mu^{(n=1)} \cdot \gamma^{(n=N)}$, where $\gamma=0.94$ is a parameter to decay learning rate. 
We trained the network for $60$ epochs for both datasets. %We trained the network for $70$ epochs for the DRIVE and $60$ epochs for IOSTAR when \textbf{default parameter number} is used and $50$ epochs for the DRIVE and $60$ epochs for IOSTAR when we halved the network parameter number.  
We optimised parameters of the proposed network with Adam algorithm,%\cite{kingma2014adam},
 with default values of $\beta_{1}$ and $\beta_{2}$, by using mini-batches of $64$ images. % --$\beta_{1}=0.9$ and  $\beta_{2}=0.999$--. 
We initialised the parameters of our network with He \textit{et al.}'s technique \cite{he2015delving}. The total number of parameters was slightly less than a million. %We keep the original number of filters for DRIVE but double it for IOSTAR, which has more variations in hue.
 
We randomly cropped $4000$ image patches from each image in the training set of the DRIVE. Of these image patches, $400$ were used for validation and the rest was allocated for training. The same process was also realised for IOSTAR dataset. The size of image patches for both datasets was set to be $96 \times 96$ pixels.

We used color images and applied channel-wise normalisation. The mean contrast for each colour channel was calculated over our training set and subtracted from training, validation and test sets.  In order to increase variety on our training sets, we applied color jittering to them, without data augmentation. Final vessel probability maps were generated by combining the output probability maps of the network for input image patches sampled with a stride of $30$ pixels at each direction for both DRIVE and IOSTAR datasets. 
 
%In addition, we applied color jittering to image patches sampled from the training set, on the fly, to make the training less sensitive to various imaging conditions and different camera characteristics. Although this operation increased the variety on contrast, blurriness and saturation of image patches on training set, it did not increase the training time. %New values for these parameters were randomly chosen from a uniform distribution of $\left[0.5, 1.5\right]$. \textbf{is is an added value over the original values or replacement of the original values ??}. 

%Gradient updates during training were made after calculating errors over mini-batches with a size of $64$ patches. Initial learning rate of $0.0008$ was exponentially reduced with $\alpha*\gamma^n$, where $\alpha$ is the learning rate, $\gamma$ is a learning rate decay parameter and $n$ is a epoch number. $\gamma$ was set to $0.94$ during experiments. \hl{The network was trained for} $60$ epochs for DRIVE, $50$ epochs for CHASE\_DB$1$ and $60$ epochs for HRF. 

%We obtained final fundus vessel maps after combining the outputs of the network with a stride of $5$ pixels for DRIVE and CHASE\_DB$1$ and with a stride of $30$ pixels for HRF. 

\subsection{Evaluation Criteria \label{sec:EvaluationCriteria}}
In order to binarize final probability maps, we found a threshold with Otsu's method.  In the binary maps, we calculated \textit{accuracy (acc)}, \textit{sensitivity (sens)}, \textit{specificity (spec)} and \textit{geometric mean (g$\_$mean)} \cite{akosa2017predictive} by considering pixels inside the FOV masks \cite{fraz2012blood}, as follows.

\resizebox{0.8\linewidth}{!}{ \begin{minipage}{\linewidth}
\begin{align} \label{eqn:EvaluationCriteria}
&acc = \frac{TN+TP}{TP+TN+FN+FP}\\
&sens= \frac{TP}{TP+FN} \\
&spec = \frac{TN}{TN+FP} \\
&g\_mean=\sqrt{sens \cdot spec}
\end{align} 
\end{minipage}
}

where $TP$, $TN$, $FN$ and $FP$ respectively denote true positives, true negatives, false negatives and false positives, which are calculated with the standard approach \cite{fraz2012blood}. A single value for each metric was computed over the complete set of test images. We also calculated the area of the Receiver Operating Characteristic curve (ROC) over final probability maps. This metric gives the discrimination capacity of the network regardless of any threshold.

%We computed a single value for each aforementioned performance metric considering all images in a test dataset. Because the ratio of vessel pixels in fundus images is far smaller than that of non-vessel pixels, \textit{Acc} may be biased by the correct estimation of non-vessel pixels. Considering this commonly encountered situation in medical image analysis, we gave more importance to G$\_$mean, MCC and F$1$ scores in our performance evaluation.

%In order to evaluate the intrinsic discrimination capability of the output of our network, we calculated the Receiver Operating Characteristic curves (ROCs) at varying threshold levels, determining $TPs$ and $FPs$ at these thresholds. The area under ROC curve ({AUC}) captures the overall discrimination capacity of the label maps \cite{fawcett2006introduction}. For consistency with other work, we report the performance metrics by considering only pixels inside the FOV masks.

\section{Results}

\subsection{Segmentation Performance}

\textbf{Table} \ref{tab:PerformanceComp} compares the segmentation performance of the proposed method with that of previous methods. According to the table, our method generates consistent performance for both DRIVE and IOSTAR datasets; though, the datasets were captured with different imaging modalities and so show different characteristics such as larger variation in hue in IOSTAR dataset.  The proposed method outperforms previous methods, including other deep leaning based studies, for both datasets with a score of $0.98$ on AUC and produces comparable or better performance on the other metrics for DRIVE dataset. For IOSTAR dataset, the proposed network outperforms other state of the art methods with a significant margin on sensitivity, with a score of $0.81$, and G-mean, with a score of $0.89$.

Indeed, the high performance of our network seems to be not due to being with large parameter count or extensive preprocessing stage but as a result of its design. The network of Liskowski and Krawiec \cite{liskowski2016segmenting} contains almost $50$ times of parameter count that our network includes. Therefore, it requires a far larger size of training dataset, which was obtained with exhaustive data augmentation \cite{liskowski2016segmenting}. Zhang \textit{et al.} \cite{zhang2016robust} and Na \textit{et al.} \cite{na2017superpixel} enhanced the appearance of vessels by applying pre-processing techniques such as homogenity correction prior to the use of their methods. Moreover, Na \textit{et al.} reported that their method could not reach to the same performance when a homogenity correction proposed by them \cite{na2017superpixel} was not used, where sensitivity and specificity respectively reduced from $0.76$ to $0.75$ and from $0.98$ to $0.92$. It should be noted that our method did not rely on such a mechanism to improve the homogenity in images; however, it still shows higher performance.

\begin{figure}[!t]
  \centering
%  \begin{subfigure}{.4\linewidth}
  {\includegraphics[scale=0.3]{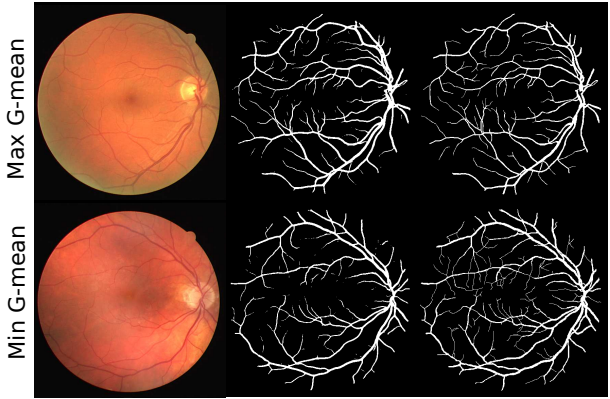}}
\caption{The segmentation performance of the proposed method on DRIVE dataset. Color fundus images are accompanied with vessel maps produced by the proposed method and ground truth segmentation masks respectively. Images from top to bottom respectively belong to $18\_test.tif$ and $07\_test.tif$. (Best viewed in colour.)}
 \label{fig:TheWorstBest_DRIVE}
 %\end{subfigure}
\end{figure}

\begin{figure}[!t]
  \centering
%  \begin{subfigure}{.4\linewidth}
  {\includegraphics[scale=0.3]{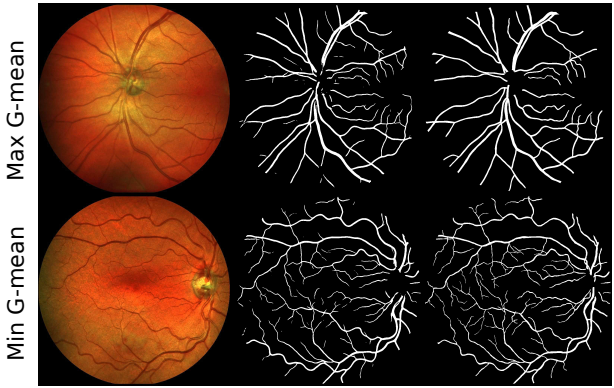}}
\caption{The segmentation performance of the proposed method on IOSTAR dataset. Color fundus images are accompanied with vessel maps produced by the proposed method and ground truth segmentation masks respectively. Images from top to bottom respectively belong to $44\_OSN.jpg$ and $34\_ODC.jpg$. (Best viewed in colour.)}
 \label{fig:TheWorstBest_IOSTAR}
 %\end{subfigure}
\end{figure}

\textbf{Fig.} \ref{fig:TheWorstBest_DRIVE} and \textbf{Fig.} \ref{fig:TheWorstBest_IOSTAR}  demonstrate segmentation maps with the maximum and minimum G-mean scores, generated by the proposed method for both DRIVE and IOSTAR. As seeing in the figures, the proposed method achieves to segment almost the complete vasculature, including many small vessels.  % We obtain the maximum sensitivity on $19^{th}$ image, with $0.93$, and the minimum sensitivity on $6^{th}$ image, with $0.75$ for DRIVE.  

%\ref{tab:PerformanceComp} also gives the performance of our method when its parameter number is reduced to almost a quarter of it. The performance of the method seems to not to be affected by the decrease in parameter number. 
%According to \ref{tab:PerformanceComp}, the proposed method manifested slightly worse performance to that of previous methods \cite{zhang2016robust,na2017superpixel} in terms of sensitivity; however, our method outperforms these methods on AUC, accuracy and specificity. Zhang \textit{et al.}'s method \cite{zhang2016robust} applied extensive preprocessing prior to the use of their method on green channel of images, which may also have an advantage of being with less variation on hue. 

%Also, Na \textit{et al.} reported that their method could not reach to the same performance when a homogenity correction proposed by them \cite{na2017superpixel} was not used, where sensitivity and specificity respectively reduced to $0.75$ and $0.92$. It should be noted that our method lacks of such a mechanism to increase homogenity in images; however, it still shows competitive performance.

%0.94496   0.96199   0.92707   0.96677   0.94671   0.83615   0.85446   0.79240 -> max sensitivity image 19
%0.84687   0.95008   0.75209   0.98261   0.85966   0.78411   0.80960   0.87663 -> min  sensitivity image 6

%  \vspace*{-0.5\baselineskip} 
\begin{table*}[t] 
\centering
\caption{Vessel segmentation performance comparison on DRIVE and IOSTAR.}
\label{tab:PerformanceComp}
\resizebox{0.7\textwidth}{!}{%
%\begin{threeparttable}
\begin{tabular}{ @{} l l  l c c c c c c c c @{}}  \toprule
Dataset&Year & Method	& &AUC	&Accuracy &Sensitivity &Specificity &G-mean\\   \cmidrule(lr){1-1}  \cmidrule(lr){2-2} \cmidrule(lr){3-3}  \cmidrule(lr){5-5} \cmidrule(lr){6-6} \cmidrule(lr){7-7} \cmidrule(lr){8-8} \cmidrule(lr){9-9} 
\multirow{8}{*}{DRIVE}

&2019&\textbf{The proposed method }&  & 0.98& 0.95& 0.81&0.98& 0.89\\
&2017&Orlando \textit{et al.} \cite{orlando2017discriminatively}&&&&0.79&0.97&0.87\\
&2016 &Liskowski and Krawiec \cite{liskowski2016segmenting} && 0.97  &  0.95     &0.75    & 0.98&0.86\\ 
&2016 & Oliveira \textit{et al.} \cite{oliveira2016unsupervised} & &0.95 & 0.95&{0.86}&0.96&{0.91}\\ 
&2015 &Li \textit{et al.}\cite{Li2015} &	& 0.97	& 0.95	& 0.76	& {0.98}&0.86\\ 
&2015&Wang \textit{et al.} \cite{wang2015hierarchical} && 0.95& {0.98}& {0.82} & 0.97&0.89\\ 
&2014 &Cheng \textit{et al.}  \cite{cheng2014discriminative} && 0.96	& 0.95	& 0.72	& 0.98&0.84	\\ 
%&2013 & Nguyen \textit{et al.} \cite{nguyen2013effective}&& & 0.94&&\\ 
&2013&Fraz \textit{et al.} \cite{fraz2013application} && & 0.94&0.73& 0.97&0.84\\ \midrule% \hdashline
\multirow{3}{*}{IOSTAR}
&2019&\textbf{The proposed method }& &0.98& 0.96&0.81& 0.98& 0.89\\
&2016& Zhang \textit{et al.} \cite{zhang2016robust} && 0.96&0.95&0.75&0.97& 0.85\\
&2017&Na \textit{et al.} \cite{na2017superpixel}&& 0.96&{0.96}&0.76&{0.98}&0.86 \\
\bottomrule								
\end{tabular}
%\begin{tablenotes}
%\item[*] The proposed network is with default parameter number, which is $995,025$.\\
%\item[**] The proposed network is with half of default number of filters, which equals to $250,933$ parameters.
% \end{tablenotes}
%\end{threeparttable}
}
\end{table*}

% Acknowledgments---Will not appear in anonymized version
%\midlacknowledgments{We thank a bunch of people.}

\subsection{Activation Map Produced at the Last Layer of Each Sub-Network}
This section will investigate roles of sub-networks on the decisions of the proposed network. Because each sub-network possess a different filter type, we expect each sub-network to behave as an expert on specific structures in input images.  %we expect activation maps produced by them to vary and to concentrate on different regions of an input image.  

\begin{figure}[!t]
 \centering
  {\includegraphics[scale=0.3]{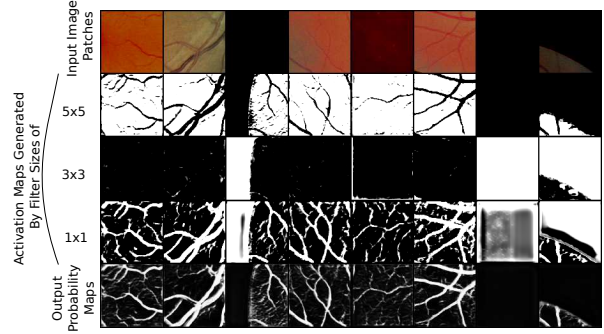}  }
  \caption{Activation maps generated at the last layers of sub-networks with filter size of $1 \times 1$, $3 \times 3$ and $5 \times 5$, demonstrated from the second row to the forth one. The first and last rows respectively show input image patches and output probability maps generated by the proposed network. Each sub-network seems to become an expert on one task: the pixel classification regarding the position of image patches inside FOV or outside FOV (the sub-network with $3 \times 3$ filters ), the classification of background pixels inside FOV (the sub-network with $5 \times 5$ filters ) and the classification of vessel pixels inside FOV (the sub-network with $1 \times 1$ filters ). Finally, the proposed network combines the activations from the sub-networks in an effective way, without exclusively outputting any of them. Images patches are randomly sampled from DRIVE dataset.  (Best viewed in color.) \label{fig:Features}}
%\vspace{-0.25cm}
\end{figure}

\textbf{Fig.} \ref{fig:Features} demonstrates activation maps at the final layers of sub-networks and output probability maps generated after combining these activation maps at the final layer of the proposed network, for eight input image patches randomly sampled from DRIVE dataset; each image patch is examined in its corresponding column in the figure. As seen in the figure, activation maps produced by filters after RELU, from top to bottom, respectively seem to focus on non-vessel pixels inside FOV, pixels outside FOV and vessel pixels inside FOV. Moreover, this specialisation of sub-networks seems to provide better description of image patches regarding the location of vessels and other structures in images.

A detailed examination of the figure reveals that vessels with low contrast, such as arteries with the central light reflection in the second input image patch and tiny vessels in the first and fifth image patches, are well recognised. The proposed network manages to assign larger probabilities to vessels despite large variation in their contrasts and thickness.

%The use of small kernel size may improve the detection of small vessels but may also lead to the loss of context, which may appear with more false positives. With the use of a range of filter size, we facilitate the network to learn global context such a pixel to be inside or outside FOV, and local information, vessel or not.    

%According to our observations in \ref{fig:Features}, we may conclude that each sub-network deals with a specific size of structures: small structures on non-vessel regions, vessel regions and large structures on non-vessel regions. Moreover, this specialisation of sub-networks seems to provide better description of image patches regarding the location of vessels and other structures in images.

%As seen in the bottom row of the figure, image background is well detected by $5 \times 5$ filter despite the sharp brightness change as a result of the presence of the optic disc and, $3 \times 3$ filter identifies vessels without compromising their continuity.

%The proposed method analyses an input image patch with three types of filter size in parallel. We expect that each filter type concentrates on a different region of an input image, depending on its filter size. For example, $1 \times 1$ filter concentrates on details and $5 \times 5$ filter deals with larger regions. In order to test our hypothesis, we evaluate activation maps obtained at layer $10$, which are eventually be combined to produce a probability map. %Features obtained at layer $10$ may give valuable insight about how the network decides on the final probability map. 

%\textbf{Although we did not use any restrictions for the filters to guide them towards the detection of details in the image background, the general location of the image background, and that of vessels, the network internally learns to classify them.  }

\section{Conclusion}
This paper presented a deep network for medical image segmentation, where the segmentation of a structure may be hindered due to the presence of pathologies, other organs and imaging related problems. The proposed network is a composite of three sub-networks, each with the same architecture but a different filter size. According to the results of experiments carried out in two fundus image datasets, DRIVE and IOSTAR, which are captured by different techniques (CCD camera and SLO), our network outperformed previous studies with a significant margin on sensitivity and G-mean for IOSTAR dataset, without using any extensive preprocessing to improve the appearance of vessels, in contrast to Zhang \textit{et al.}'s and Na \textit{et al.}'s methods. Also, our network showed better or comparable performance on DRIVE when compared with that of previous methods. 

We show that each sub-network attuned to specific tasks such as the identification of pixels outside FOV and the classification of vessel pixels from others inside FOV. One may find similarities between the output of our sub-networks and attention maps, which is designed to pay attention to such regions of an image that may facilitate the detection or segmentation of an object of interest \cite{chen2016attention}. Despite our network consisting of three sub-networks, its total parameter number is far smaller than that of other networks producing similar performance\cite{liskowski2016segmenting}. This allows the proposed network to be easily applied for image segmentation tasks with the limited amount of labelled data. 

 %In the present work, we apply the spirit of inception modules in a global scale. In the communication of immediate neighbour layers or layers linked with skip connections, we only allow the transmission of information between the same size of filters. \ref{fig:Method}(a) illustrates the general structure of the proposed architecture. In other words, an input image is analysed with the same size of filters through the network as it is the case in U-net. However, training is realised as if we run three parallel sub-networks, each with a different size of filters. To be more precise, this network may be viewed as a composite of $3$ sub-networks, each with a single filter size, connected in parallel as demonstrated in \ref{fig:Method}(b).  

%Although we only provide two classes for the segmentation task at hand, vessel or not, the proposed network found an additional class, which shows  details in the image background. \textbf{we observed that reducing the filter number of the penultimate layer of the U-net to 3 did not lead to the same grouping of activation maps we obtained by the proposed method ???? }

%We will investigate if $1 \times 1$ filter in the penultimate layer may be useful for detection of any anomalies or diseases in fundus images. We will also examine its validity on other medical images. 

%\midlacknowledgments{The author thanks to Anil A. Bharath for his valuable feedback on the paper.}

\bibliographystyle{plain}
\bibliography{midl-samplepaperEleco}

%\appendix
%\section{Appendix}

\end{document}